\documentclass[final,3p,times,twocolumn]{elsarticle}
\usepackage{xcolor}

\usepackage{amssymb}
\usepackage{adjustbox}
\usepackage{subfigure}
\journal{Journal of High Energy Astrophysics}

\begin{document}

\begin{frontmatter}



\title{Accretion disk wind during the outburst of the stellar-mass black hole MAXI J1348-630 }

\author[label1,label2]{Hanji Wu}
\author[label1,label2]{Wei Wang}
\ead{wangwei2017@whu.edu.cn}
\author[label1,label2]{Na Sai}

\address[label1]{Department of Astronomy, School of Physics and Technology, Wuhan University, Wuhan 430072, China}
\address[label2]{WHU-NAOC Joint Center for Astronomy, Wuhan University, Wuhan 430072, China}

\begin{abstract}
We analyzed two observations of the low-mass black hole X-ray binary MAXI J1348-630 from Nuclear Spectroscopic Telescope Array (NuSTAR) during low hard state and hard intermediate state in the 2019 outburst. The reflection components are found in the X-ray spectra, and the spectral fittings give an inclination angle of $\sim 25^\circ-35^\circ$ for the binary system, and there is an absorption line around $\sim 7$ keV coming from highly ionized iron. The photoionization code XSTAR is used to fit the absorption line, which is attributed to outflows with a velocity of $\sim 10^{4}\rm km\ s^{-1}$ along our line of sight and the column density reaching $10^{23}\rm cm^{-2}$ in low hard and hard intermediate states. The physical mechanism launching fast disk winds from the black hole accretion system is still uncertain. These observations strongly support magnetic launching as the dominant mechanism which drives the high velocity, high ionization winds from the inner accretion disk region in hard and hard intermediate states of MAXI J1348-630. 
\end{abstract}




\end{frontmatter}



\section{Introduction}
The primary radiation characteristic of a black hole X-ray binary (BHXRB) can be modeled by multi-temperature black body spectra \citep{shakura1973black,novikov1973astrophysics}, while at the hard X-ray bands, the BHXRB spectra have a power-law component that results from the accretion disk radiation scattered by the high-temperature plasma known as corona \citep{sunyaev1980comptonization,sunyaev1985comptonization}. However, a power-law component fails to model the iron emission lines at 6-7 keV and a hump at over 20 keV, which is modeled well by Comptonized photons reflected by the disk \citep{fabian2016innermost}. Furthermore, the iron emission line would be reshaped as a relativistically broadened and asymmetric iron line \citep{fabian1989x,reynolds2003fluorescent} and reveal the geometry of the corona and the motion of the corona after adding a boosting ionized component \citep{you2021insight}. The Compton and reflection components are commonly found in the low hard state while in the high soft state these two components would be suppressed \citep{remillard2006x,done2007modelling}.

In X-ray spectra of low mass X-ray binary (LMXB) systems, it is common to find the ionized absorption component, which is regarded as the disk winds normally \cite{lee2002high,miller2015powerful,diaz2016accretion,king2012extreme,neilsen2018persistent}. In X-ray bands, the most common absorption lines are the Fe \uppercase\expandafter{\romannumeral25} and the Fe \uppercase\expandafter{\romannumeral26}. The accretion disk loses the mass which could be matched with the mass accretion rate by wind \citep{ueda2009grs,ponti2012ubiquitous}. There are three models to launch outflows in accreting systems, namely, the thermal pressure \citep{begelman1983compton,woods1996x,done2018thermal}, radiation pressure \citep{proga2002role,higginbottom2015coronae} and magnetic pressure \citep{miller2006magnetic,fukumura2017magnetic,wang2022magnetically}. However, distinguishing the mechanisms that outflows are launched is not easy, because the launch position and the wind density are difficult to determine. In observation, the wind and jet are quite mutually exclusive \citep{neilsen2009accretion}. Furthermore, the disk winds are more common in the high soft state \citep{miller2008accretion,ponti2012ubiquitous} than in the low hard state. However, the reason for this is still quite uncertain if the disk winds are not launched in the low hard state \citep{neilsen2012hybrid} or winds are launched, but due to the high ionization \citep{shidatsu2019application} or thermally unstable \citep{chakravorty2013effects,petrucci2021expected}, so that the absorption features are difficult to find in the low hard state. 

MAXI J1348-630 was discovered by the Gas Slit Camera (GSC) onboard Monitor of All-sky X-ray Image on Jan 26 2019 \citep{matsuoka2009maxi,yatabe2019maxi}, and classified as a black hole X-ray binary (BHXRB) by the mass estimate and spectral features \citep{tominaga2020discovery}. The Swift/XRT \citep{tominaga2020discovery} data and NICER data \citep{zhang2020nicer} show that this source obeyed the "Q"-shape feature during the 2019 outbursts. SRG/eROSITA data revealed a giant dust scattering ring around this source, and with the joint data from XMM-Newton, MAXI, and Gaia, estimated the distance of MAXI J1348-630 at 3.39 kpc and the black hole mass of 11 $\pm 2 M_{\odot}$ \citep{lamer2021giant}. By the Australian Square Kilometre Array Pathﬁnder (ASKAP) and MeerKAT Data, the H \uppercase\expandafter{\romannumeral1} absorption spectra of this source determined the distance as $2.2^{+0.5}_{-0.6}$ kpc \citep{chauhan2021measuring}. Furthermore, both type-B quasi-periodic oscillations (QPO) \citep{belloni2020time} and the type-C QPOs were found in this BHXRB \citep{alabarta2022variability}. The timing analysis indicates that there is a time lag in the energy band of 0.5-80 keV \citep{jithesh2021broad} as well as the evidence of the time lag between the disk component and the corona radiation has been discovered \citep{weng2021time}. Disk wind signatures have been found in the infrared and optical bands \citep{panizo2022discovery}. 

The Nuclear Spectroscopic Telescope Array (NuSTAR, \citep{harrison2013nuclear}) has performed several observations of MAXI J1348-630 during the 2019 outburst in low hard state, hard intermediate state, soft state, which shows evidence of a high density disk reflection component \citep{chakraborty2021nustar} and constrains the physical properties of the central compact object \citep{jia2022detailed,kumar2022estimation}. The Insight-HXMT has the advantage of a high energy band ($>$ 20 keV) and observed the entire outburst from January- July 2019 \citep{chen2019insight}. This source has shown the state transition like the other BHXRBs \citep{homan2005evolution,fender2004towards} from hard to soft states (from MJD 58517 to MJD 58530) and a mini-burst after the end of the outburst (around MJD 58600) \citep{tominaga2020discovery}. 
 

With the high spectral resolution of NuSTAR in X-ray bands, the absorption features around 7 keV were found in both of the hard state and hard intermediate state, which could be the Fe \uppercase\expandafter{\romannumeral26} absorption. Here, we try to constrain the physical properties of the ultrafast winds in MAXI J1348-630 by using the photoionization code XSTAR. This paper is organized as follows: the observation and data reduction in section \ref{data reduction}, data analysis and photoionization modeling results in section \ref{data analysis}, conclusion and discussion in section \ref{conclusion}.

\section{Observations and Data Reduction}\label{data reduction}
\subsection{NuSTAR}
The outburst of MAXI J1348-630 in January-June 2019 was observed by NuSTAR \citep{harrison2013nuclear} which covered 9 times of observations in 8 different epochs, from low hard state to soft state as well as the following mini-burst. The spectral analysis finds the absorption feature in two observations shown in Table \ref{nudata}, then we concentrate on the two observations by modeling the absorption lines. The NuSTAR data\footnote{https://heasarc.gsfc.nasa.gov/FTP/nustar/data/} are processed by NuSTARDAS v2.1.1\footnote{https://nustarsoc.caltech.edu/NuSTAR\_Public/NuSTAROperationSite/\\ software\_calibration.php} released on 2021-03-18 including NuSTAR pipelines with version 0.4.9 and nuproducts with version 0.3.3 and the NuSTAR CALDB v20220328\footnote{https://heasarc.gsfc.nasa.gov/docs/heasarc/caldb/\\ caldb\_supported\_missions.html}, which will correct the FPMA spectral excess, as compared to FPMB, in low energy ($\textless$7keV) resulted by a rip in the Multi Layer Insulation (MLI) on the NuSTAR Optics Module A (OMA) \citep{madsen2020nustar}. The issue flags of two observations (Obs ID:80402315008/80402315008) indicate that there are background flares due to enhanced solar activity and we set saacalc = 2, saamode = optimized, tentacle = no to filter them in NUPIPELINE. 

Furthermore, the flux of MAXI J1348-630 was over 4 Crab unit at the outburst peak, which could generate spurious triggers in the NuSTAR cleaned event files, so we modified the parameter 'statusexpr' in NUPIPELINE, statusexpr="STATUS==b0000xxx00xxxx000" to remove spurious triggers, suggested by the NuSTAR Analysis Guide\footnote{https://heasarc.gsfc.nasa.gov/docs/nustar/analysis/}. We use FTOOLS: ftgrouppha to opt the spectra by the parameter "grouptype = optsnmin, groupscale = 19" \citep{kaastra2016optimal}. For joint analysis of the spectra from FPMA and FPMB, a cross-normalization constant (a constant model from XSPEC) is used to measure the difference between them as FPMA is unity. After \cite{madsen2020nustar} correct the rip of MLI, the fitting energy range is considered from 3 keV to 75 keV, besides, there are instrumental features in 11.5 keV and 26.5 keV in bright sources reported by \cite{xu2018reflection}, thus, the energy ranges 11 keV - 12 keV and 26 keV - 28 keV are excepted in the analysis.
        \begin{table}
		\centering
		\caption{The observation dates and exposure time of two NuSTAR observations. The observation date is in the form of years-months-days.}
		\label{nudata}
		\begin{tabular}{lcr} 
			\hline
			Observation ID & DATE & Exposure(s) \\
			\hline
			80402315004 & 2019-02-01 & 736 \\
			80402315006 & 2019-02-06 & 4520 \\
			\hline
		\end{tabular}
	    \end{table}
		\begin{table*}
		\centering
		\caption{Spectral fitting results for Observation 404, $\xi_{wind}$ is ionization parameter of the wind which is calculated by the xstar model, and $\xi_{disk}$ is the ionization parameter for the inner accretion disk which is calculated by the relxill(Cp) model.}
		\label{404p}
		\begin{adjustbox}{center}
		\begin{tabular}{lccccr} 
			\hline
			components & parameter & model 1 & model 2 & model 3 & model 4 \\
			\hline
			xstar & column density ($10^{22}$ $\rm cm^{-2}$) & \dots &\dots & $23.44^{+23.65}_{-14.78}$ & $14.43^{+27.86}_{-7.72}$  \\
			       & log$\xi_{wind}$ $\rm (erg\ cm\ s^{-1})$ & \dots &\dots & $3.40^{+2.47}_{-0.50}$ & $3.17^{+2.60}_{-0.25}$ \\
			       & $v$ $(km\ s^{-1})$ & \dots &\dots & $12124^{+5237}_{-3752}$ & $12540^{+4525}_{-3627}$ \\
			\hline
			TBabs & $N_{\rm H}$ $(\times 10^{22}\rm cm^{-2})$ & $0.86^{\dag}$ & $0.86^{\dag}$ & $0.86^{\dag}$ & $0.86^{\dag}$ \\
			
			Diskbb & $kT_{in}$ $(\rm keV)$ & $0.68^{+0.03}_{-0.12}$ & $0.63^{+0.06}_{-0.10}$ & $0.76^{+0.05}_{-0.12}$ & $0.62^{+0.06}_{-0.07}$ \\
			       & norm & $3704^{+7649}_{-754}$ & $4163^{+8094}_{-1817}$ & $2173^{+2957}_{-581}$ & $4142^{+4587}_{-1811}$ \\
			\hline
			relxill(Cp) & $R_{in}$ $(\rm r_{g})$ & $5.53^{+0.68}_{-2.53}$ & $8.93^{+3.62}_{-4.33}$ & $9.38^{+4.98}_{-7.27}$ & $14.84^{+30.72}_{-4.29}$ \\
			 & i (deg) & $26.24^{+6.33}_{-1.11}$ & $25.73^{+5.40}_{-0.62}$ & $41.98^{+9.48}_{-27.92}$ & $44.33^{+14.03}_{-13.27}$ \\
			 & $\Gamma$ & $1.49^{+0.01}_{-0.05}$ & $1.68^{+0.01}_{-0.02}$ & $1.47^{+0.06}_{-0.03}$ & $1.68^{+0.02}_{-0.01}$ \\
			 & log$\xi_{disk}$ $\rm (erg\ cm\ s^{-1})$ & $3.61^{+0.57}_{-0.02}$ & $3.70^{+0.28}_{-0.28}$ & $3.49^{+0.56}_{-0.23}$ & $3.62^{+0.19}_{-0.30}$ \\
			 & $\rm A_{Fe}$ $(A_{Fe,\odot})$ & $4.99^{+4.83}_{-0.48}$ & $3.99^{+3.62}_{-1.95}$ & $4.34^{+5.21}_{-1.79}$ & $2.76^{+1.90}_{-1.49}$ \\
			 & $R_{ref}$ & $0.18^{+0.25}_{-0.01}$ & $0.13^{+0.07}_{-0.02}$ & $0.24^{+0.11}_{-0.07}$ & $0.24^{+0.14}_{-0.10}$ \\
			 & norm & $0.22^{+0.01}_{-0.04}$ & $0.21^{+0.01}_{-0.01}$ & $0.20^{+0.01}_{-0.01}$ & $0.20^{+0.01}_{-0.01}$ \\
			 & Ecut (keV) & $97.14^{+35.12}_{-6.21}$ & \dots & $81.86^{+25.682}_{-7.18}$ & \dots \\
			 & $kT_{e}$ $\rm (keV)$ & \dots &$21.23^{+2.85}_{-1.15}$ & \dots & $22.56^{+3.68}_{-2.14}$ \\
			 & log$N$ $\rm (cm^{-3})$ & \dots & $16.55^{+2.94}_{-1.42}$ & \dots & $16.98^{+2.07}_{-0.21}$ \\
			 & $C_{FPMB}$ & $0.99^{+0.003}_{-0.002}$ & $0.99^{+0.003}_{-0.002}$ & $0.99^{+0.003}_{-0.002}$ & $0.99^{+0.002}_{-0.003}$ \\
			 & $\chi^{2}/\nu$ & $956.7/905$ & $943.0/905$ & $925.6/905$ & $917.93/905$ \\
			 & $\chi^{2}_{\nu}$ & $1.06$ & $1.04$ & $1.02$ & $1.01$ \\
			\hline
		\end{tabular}
		\end{adjustbox}
	    \end{table*}
	    
\subsection{Insight-HXMT}
The Insight-HXMT is the first Chinese X-ray observatory, launched on June 15 2017, which provides over 50 days observation \citep{zhang2020overview} from Jan 27 2019 to Jul 29 2019. The data is recorded by the three telescopes: the Low Energy (LE) telescopes with 384 $\rm cm^{2}$ effective area from 1 keV to 15 keV \citep{chen2020low}; the Medium Energy (ME) telescopes with 952 $\rm cm^{2}$ effective area from 5 keV to 30 keV \citep{cao2020medium}; the High Energy (HE) telescope with 5100 $\rm cm^{2}$ from 20 keV to 250 keV \citep{liu2020high}. The good time intervals (GTI) are screened by the recommended criteria: the geomagnetic cutoff rigidity $\textgreater$ 8 GeV, the elevation angle $\textgreater$ 10 degrees, the pointing offset angle $\textless$ 0.1 and at least 300 s away from the South Atlantic Anomaly (SAA). The spectrum is extracted by the Insight-HXMT Data Analysis software (HXMTDAS) v2.04\footnote{http://hxmtweb.ihep.ac.cn/software.jhtml}. For joint analysis of the LE, ME, HE observations, we construct the data from LE telescope of 1-10 keV, ME telescope of 10-30 keV, HE telescope of 30-100 keV. Due to the lower spectral resolution at 6 -- 7 keV\footnote{http://hxmtweb.ihep.ac.cn/AboutHxmt.jhtml} than NuSTAR, the Insight-HXMT spectrum cannot constrain the absorption features, thus in this paper, we only use the Insight-HXMT data producing the light curves of three detectors to show the state transitions.

\section{Data Analysis}\label{data analysis}
We use XSPEC v12.12.1\footnote{https://heasarc.gsfc.nasa.gov/xanadu/xspec/} to fit the spectra. For fitting NuSTAR spectra of FPMA and FPMB simultaneously, we implemented a cross-normalisation constant in FPMA and FPMB (freezing the FPMA constant as standard, FPMB constant as a variable). In all of the models, the galactic neutral hydrogen column density is fixed to be 8.6$\times 10^{21} \rm cm^{-2}$ at the composition TBabs. The uncertainties were settled at 90\% confidence level for every parameter. As discussed in Section 2.1, we chose two ObsIDs 80402315004 (low hard state) and 80402315006 (hard intermediate state) for detailed spectral analysis  (hereafter two ObsIDs abbreviated as 404 and 406, respectively). 
        
\subsection{State evolution}
\begin{figure*}
    \centering
    \includegraphics[scale=1]{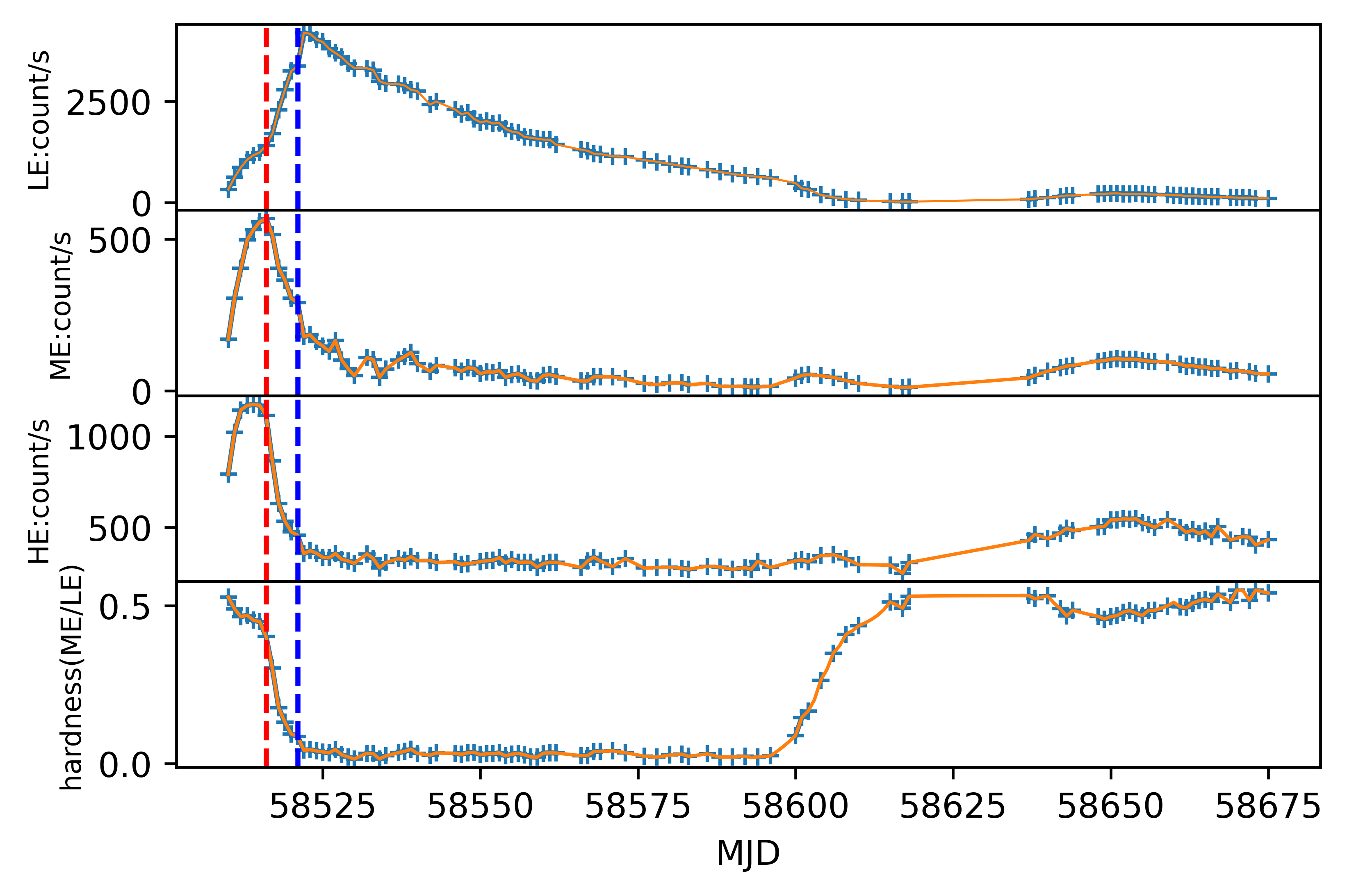}
    \caption{Light curves of MAXI J1348-630 in units of count rates by Insight-HXMT. The top panel illustrates the light curve recorded by the LE telescopes, the next one for ME, the third one for HE, the bottom for the hardness which is the count ratio between ME and LE. The red line represents the NuSTAR ObsID 80402315004, the blue line for the ObsID 80402315006. }
    \label{lightcurve}
\end{figure*}

\begin{figure}
    \centering
    \includegraphics[scale=0.5]{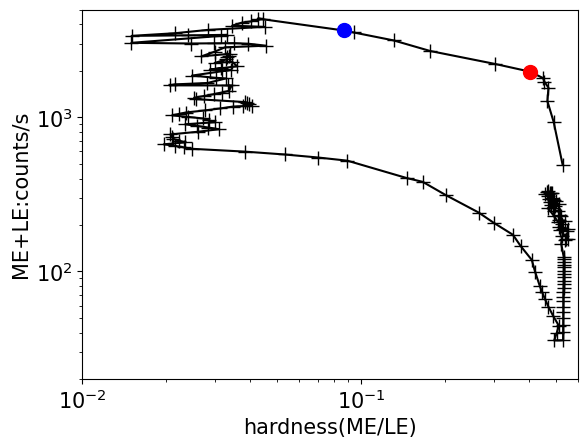}
    \caption{The HID of MAXI J1348-630 in units of count rates observed by Insight-HXMT. The hardness (the count rate ratio between ME and LE) verses the total counts rate recorded by ME and LE telescopes. The red dot represents the NuSTAR ObsID 80402315004, the blue dot for the ObsID 80402315006.}
    \label{q}
\end{figure}

The Insight-HXMT observations of this outburst from MJD 58510 to MJD 58693 with the light curves of three telescopes and hardness defined as the count rate between ME and LE are presented in Fig. \ref{lightcurve}. The outburst of MAXI J1348-630 followed the classical "Q" pattern in the hardness intensity diagram (HID) which is also illustrated in Fig. \ref{q}. From MJD 58510 to 58516, the ME and HE count rates increased with time, with the hardness around 0.5, the BHXRB was in low hard state (LHS), and from MJD 58516 to 58522, the rates of ME and HE decayed, and LE rate increased very rapidly, the hardness decreased to below $0.1$, the source in hard intermediate state (HIS). With the LE rate reaching a peak and slowly decaying, hardness around $\sim 2\%$, the source was located in soft intermediate state (SIS) and soft state (SS). From MJD 58600, the source turned into the quiescence with hardness back to $\sim 0.5$, while after one month, a mini-outburst appeared with the hardness at constant of $\sim 0.5$ lasting about 20 days. Two NuSTAR observations for the spectral analysis are marked in the red (ObsID 404) in LHS and the blue (ObsID 406) in HIS (see Figs. \ref{lightcurve} and \ref{q}). 


\subsection{The spectral fittings: continuum and absorption features}

We will analyze the spectral properties of two NuSTAR observations in LHS and HIS, when the thermal disk component is weak, and a Comptonization component from corona, usually modeled by a power-law spectrum with a high-energy cutoff would dominate the spectrum. In addition, the Comptonized photons back onto the disk may produce a reflection spectrum which has been reported in the previous work on MAXI J1348-630 \citep{chakraborty2021nustar,jia2022detailed}. In this section, we have taken the following models to fit the spectra, especially describing the strong reflection component in the hard state.

\subsubsection{Model 1: the reflection model}\label{model 1}
The broad iron line and the Compton hump indicate that there is a reflection component from the optical thick accretion disk near the black hole which is a typical feature in black hole X-ray binary spectra \citep{jiang2020nustar,walton2016soft,miller2013nustar,fabian1989x}. For fitting these, the self-consistent relativistic disc reflection model from RELXILL model suite (relxill v 2.0 \cite{garcia2015estimating,dauser2014role}) has been used. The emissivity index is fixed at $q_{1}=q_{2}=3$ as the recommendation \citep{fabian1989x}. We start with the basic model relxill and the constitution of Model 1 is:
$$
    tbabs\times(diskbb+relxill) ({\rm Model 1}).
$$

Due to the correlation between the black hole spin parameter ($a$) and the inner radius of the accretion disk ($R_{in}$), the black hole spin is fixed as the maximum ($a=0.998$) and let the inner radius ($R_{in}$) be free in the reflection component model and the outer radius of the accretion disk fixed as 400 $r_{g}$. The model gave an acceptable fit $(\chi^{2}/\nu = 1398.76/1194)$ and resolved the 6.4 keV broad iron line and the Compton hump (more details of fitting results shown in the model 1 column of Table \ref{404p} \& \ref{406p}) and the spectra and residuals of the Model 1 to fit the spectrum for ObsID 406 are shown in the \ref{406M1cb}. 

The 7 keV Fe \uppercase\expandafter{\romannumeral26} absorption feature appeared in the residuals, and this feature was also indicated in the previous work \citep{chakraborty2021nustar}.

\subsubsection{Model 2: the Comptonized reflection model}\label{model 2}
To confirm the absorption line, we test a more physical model in the RELXILL model suite, relxillCp which is different from relxill with incident spectra made by nthcomp model \citep{zdziarski1996broad,zycki19991989}. The constitution of Model 2 is:
$$
    tbabs\times(diskbb+relxillCp) ({\rm Model 2}).
$$
The absorption line still exists in the residuals as presented in the right panel of Fig. \ref{406M2cb} and the fitting result parameters of the Model 2 fitting are shown in the model 2 column of Table \ref{404p} \& \ref{406p}.
        \begin{table*}
		\centering
		\caption{Spectral fitting results for Observation 406, $\xi_{wind}$ is ionization parameter for the wind which is calculated by the xstar model, and $\xi_{disk}$ is the ionization parameter for the inner accretion disk which is calculated by the relxill(Cp) model.}
		\label{406p}
		\begin{adjustbox}{center}
		\begin{tabular}{lccccr} 
			\hline
			components & parameter & model 1 & model 2 & model 3 & model 4 \\
			\hline
			xstar & column density ($10^{22}\rm cm^{-2}$) & \dots &\dots & $2.60^{+4.05}_{-1.43}$ & $1.85^{+27.68}_{-0.06}$  \\
			       & log$\xi_{wind}$ $\rm (erg\ cm\ s^{-1})$ & \dots &\dots & $2.81^{+0.13}_{-0.22}$ & $2.73^{+0.17}_{-0.31}$ \\
			       & $v$ $\rm (km\ s^{-1})$ & \dots &\dots & $18846^{+1247}_{-4252}$ & $15670^{+3253}_{-4208}$ \\
			\hline
			TBabs & $N_{\rm H}$ $(\times 10^{22}\rm cm^{-2})$ & $0.86^{\dag}$ & $0.86^{\dag}$ & $0.86^{\dag}$ & $0.86^{\dag}$ \\
			
			Diskbb & $kT_{in}$ $(\rm keV)$ & $0.66^{+0.02}_{-0.005}$ & $0.68^{+0.01}_{-0.01}$ & $0.67^{+0.02}_{-0.01}$ & $0.69^{+0.01}_{-0.04}$ \\
			       & norm & $24970^{+943.5}_{-3735.5}$ & $23673^{+1238}_{-1580}$ & $25587^{+3742.60}_{-836.10}$ & $21530^{+9195.67}_{-2112.44}$ \\
			\hline
			relxill(Cp) & $R_{in}$ $(\rm r_{g})$ & $1.90^{+2.49}_{-0.19}$ & $2.10^{+0.38}_{-0.18}$ & $1.74^{+3.66}_{-0.39}$ & $1.89^{+0.06}_{-0.22}$ \\
			 & i (deg) & $35.47^{+0.47}_{-5.94}$ & $28.93^{+0.66}_{-6.69}$ & $37.66^{+1.95}_{-4.09}$ & $29.76^{+3.02}_{-3.96}$ \\
			 & $\Gamma$ & $2.28^{+0.01}_{-0.03}$ & $2.18^{+0.01}_{-0.05}$ & $2.20^{+0.05}_{-0.003}$ & $2.14^{+0.02}_{-0.01}$ \\
			 & log$\xi_{disk}$ $\rm (erg\ cm\ s^{-1})$ & $4.26^{+0.11}_{-0.03}$ & $4.54^{+0.13}_{-0.11}$ & $4.55^{+0.02}_{-0.024}$ & $3.81^{+0.26}_{-0.19}$ \\
			 & $\rm A_{Fe}$ $\rm(A_{Fe,\odot})$ & $5.05^{+1.12}_{-0.39}$ & $8.00^{+1.69}_{-0.51}$ & $9.09^{+0.49}_{-5.79}$ & $3.45^{+1.89}_{-0.60}$ \\
			 & $R_{ref}$ & $0.58^{+0.05}_{-0.12}$ & $0.80^{+0.08}_{-0.16}$ & $0.83^{+0.49}_{-0.12}$ & $0.93^{+0.22}_{-0.18}$ \\
			 & norm & $0.25^{+0.01}_{-0.03}$ & $0.15^{+0.02}_{-0.02}$ & $0.18^{+0.03}_{-0.05}$ & $0.13^{+0.02}_{-0.01}$ \\
			 & Ecut (keV) & $304.2^{+52.23}_{-54.89}$ & \dots & $233.99^{+88.54}_{-6.17}$ & \dots \\
			 & $kT_{e}$ $\rm (keV)$ & \dots &$>400$ & \dots & $392.5^{+0.13}_{-249.52}$ \\
			 & log$N$ $\rm (cm^{-3})$ & \dots & $18.50^{+0.27}_{-1.26}$ & \dots & $18.52^{+0.44}_{-0.73}$ \\
			 & $C_{FPMB}$ & $0.98^{+0.001}_{-0.001}$ & $0.98^{+0.001}_{-0.001}$ & $0.98^{+0.001}_{-0.001}$ & $0.98^{+0.001}_{-0.001}$ \\
			 & $\chi^{2}/\nu$ & $1437/1194$ & $1449/1194$ & $1357.26/1194$ & $1390.86/1194$ \\
			 & $\chi^{2}_{\nu}$ & $1.20$ & $1.21$ & $1.14$ & $1.16$ \\
			\hline
		\end{tabular}
		\end{adjustbox}
	    \end{table*}

The new version 2.0 of the RELXILL model suite updates the model relxillCp by adding the parameter $\log N$ which is the log of the disk density. The disk density is about 1000 times higher than the standard model $10^{15}\rm cm^{-3}$ otherwise the abundance of iron could be higher (compare Model 1 and Model 2 about $A_{Fe}$).


\subsubsection{Model 3 \& 4: the reflection model with absorption lines}
The above reflection models have shown the absorption features around 7 keV in both two observed spectra, so we at first add a Gaussian absorption component to check the absorption line properties, finding the absorption line energy at $\sim 7.2$ keV and the line width of about 250 eV for 404 and 150 eV for 406. This absorption feature is generally attributed to the winds from the accretion system, and the line width should be due to the turbulent velocity in the wind.

\begin{figure*}
\centering
\subfigure[The spectra and residuals of Model 1.]
{
    \includegraphics[scale=0.5]{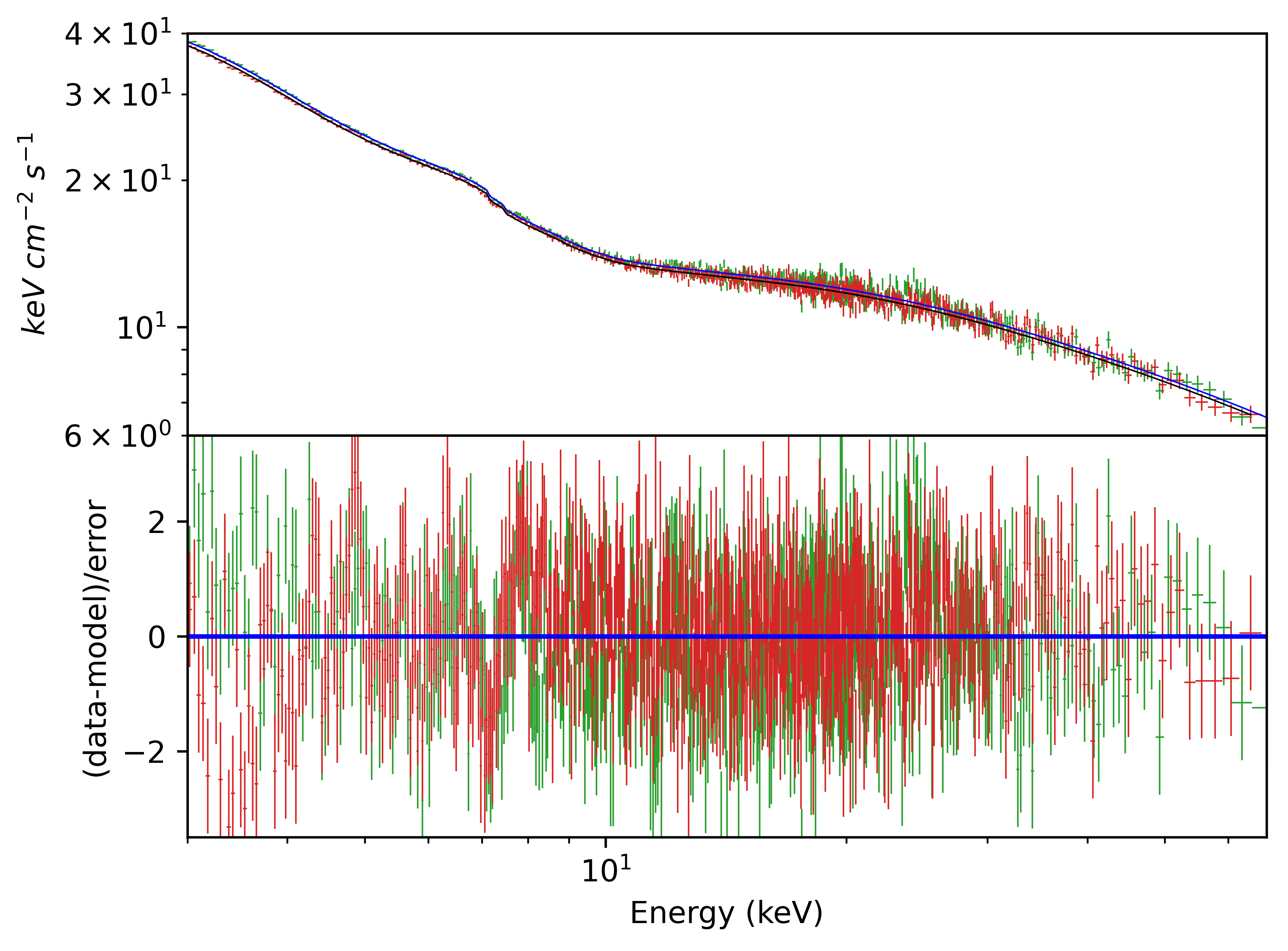}
    \label{406M1cb}
    }
\quad
\subfigure[The spectra and residuals of Model 2.]
{
    \includegraphics[scale=0.5]{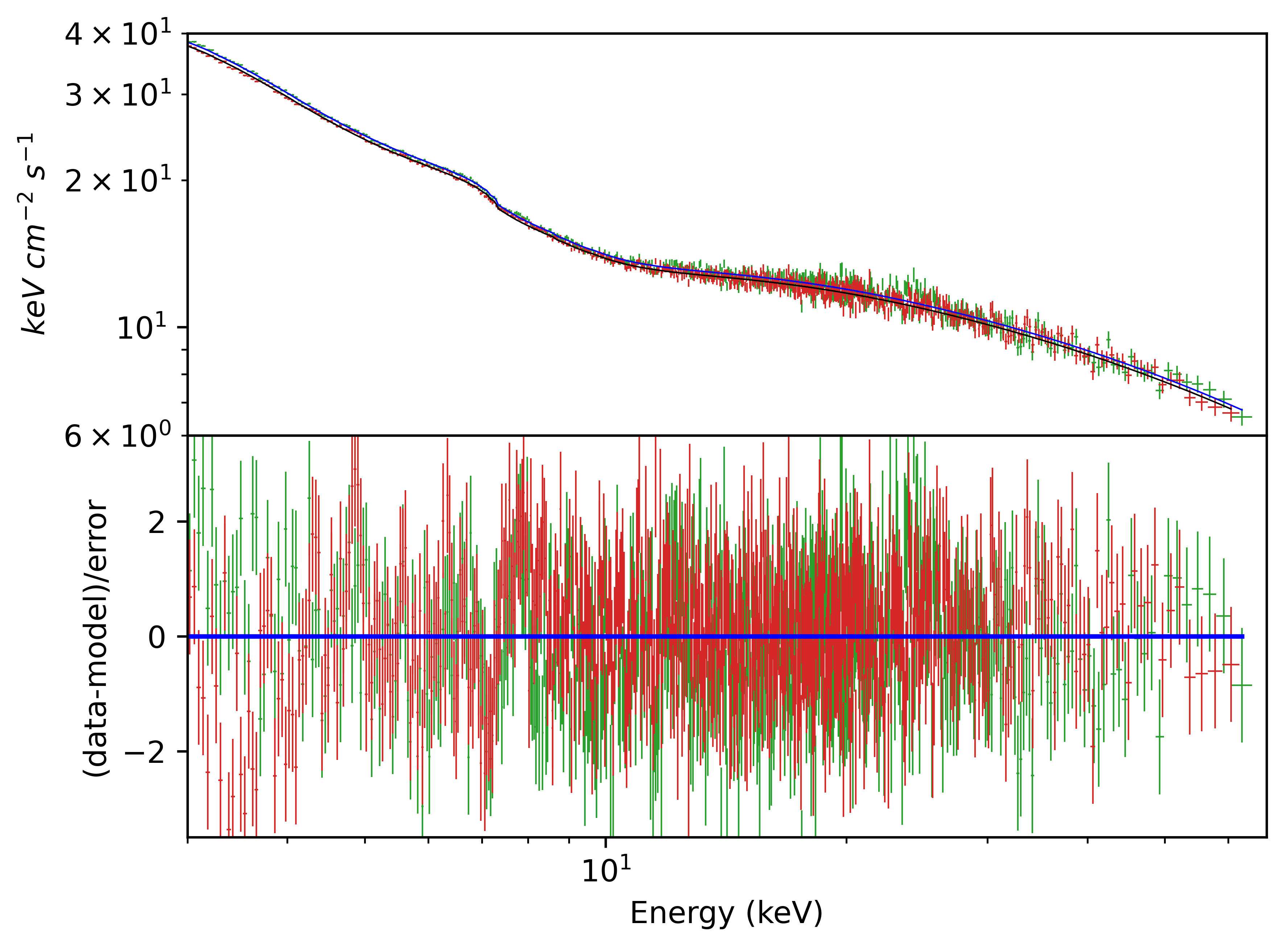}
    \label{406M2cb}
 }
\quad
\subfigure[The spectra and residuals of the Gauss absorption lines component test model.]
{
    \includegraphics[scale=0.5]{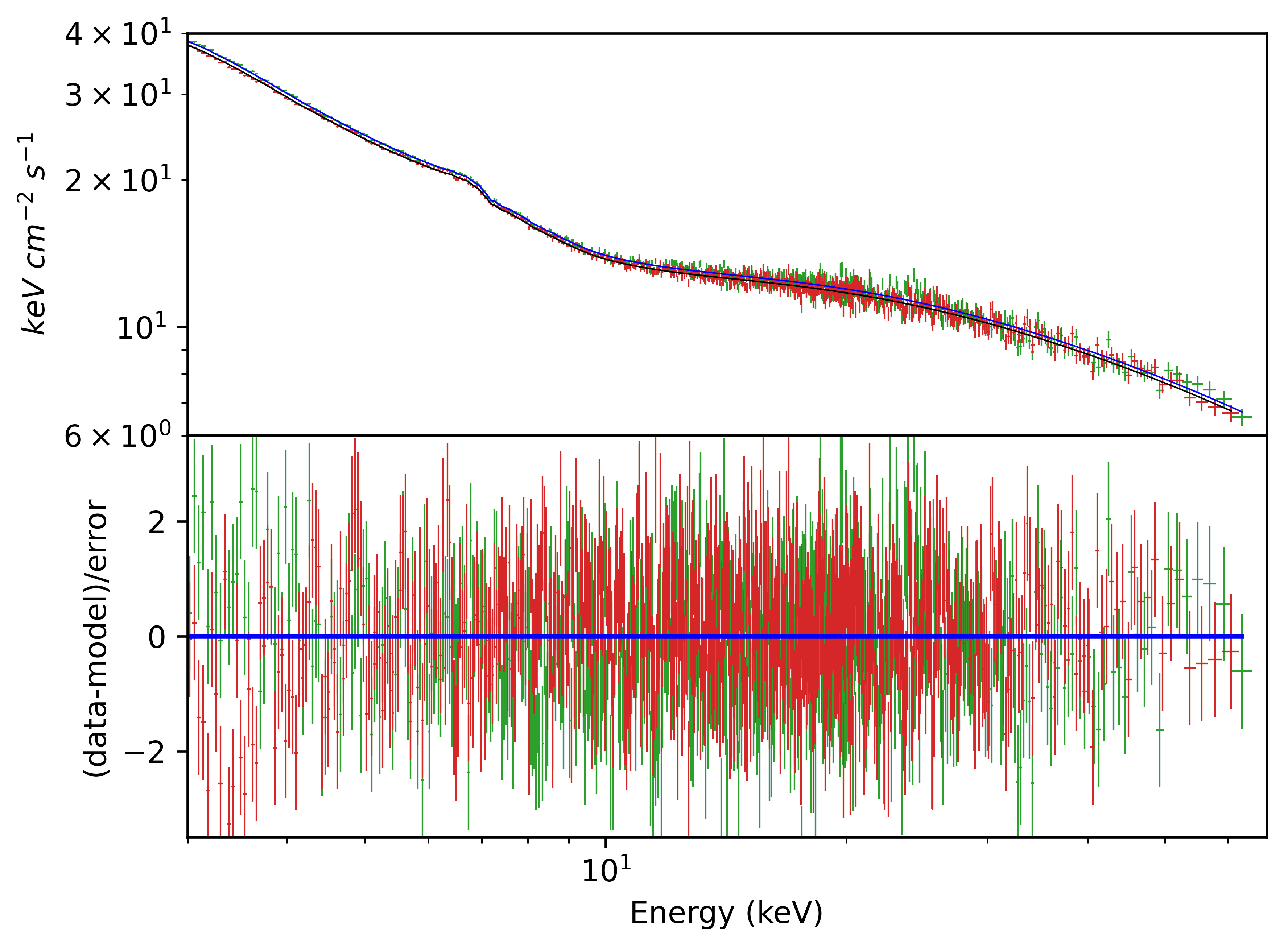}
    \label{406CpHDg}
}
\quad
\subfigure[The spectra and residuals of Model 4.]
{
    \includegraphics[scale=0.5]{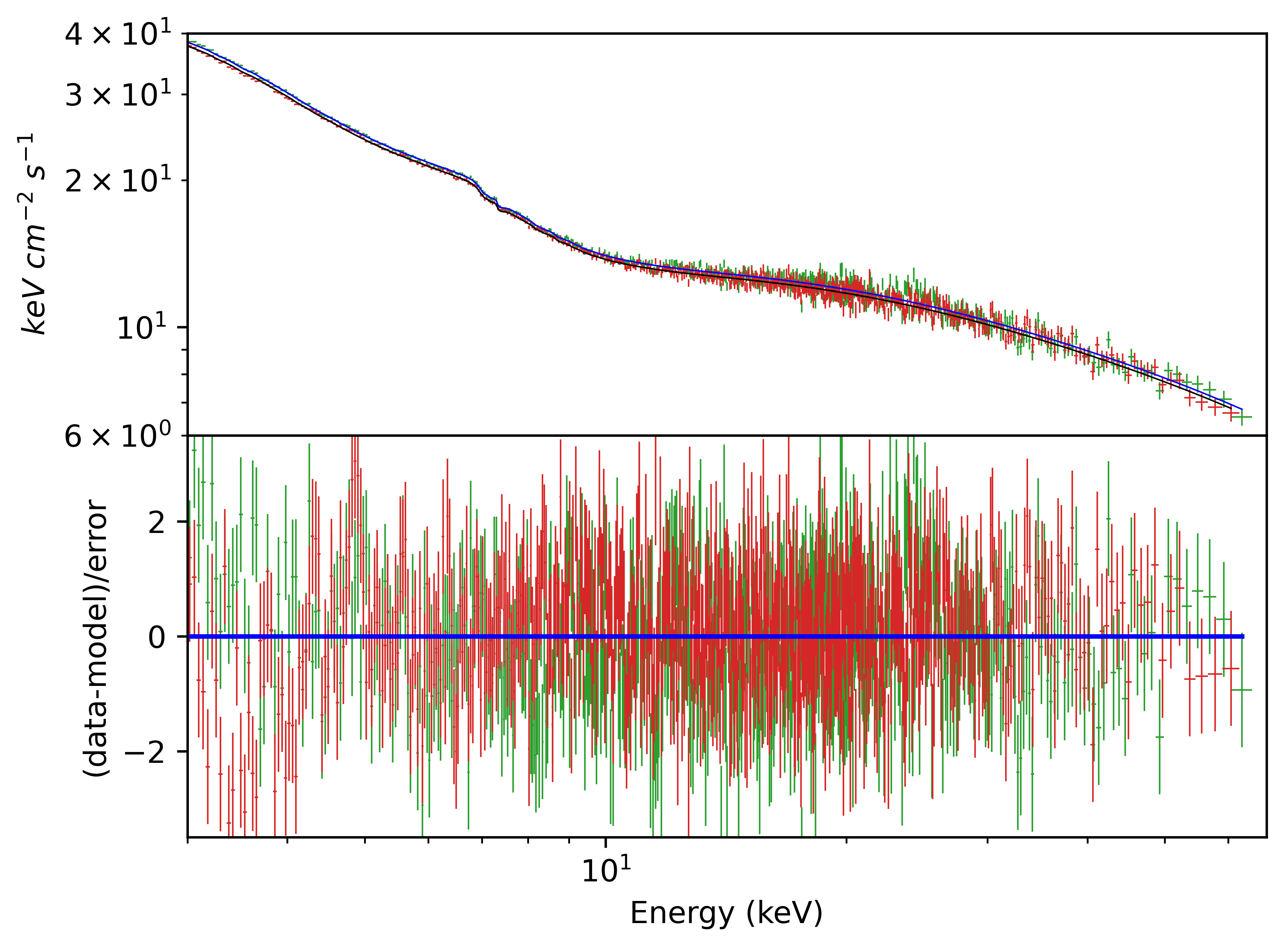}
    \label{406M4cb}
}
\caption{Four pictures show the spectra for ObsID 406 fitted with all four different models, the green and the red data points correspond to the FPMA and FPMB, respectively. }
\end{figure*}

\begin{figure*}
     \centering
     \includegraphics[scale=0.4]{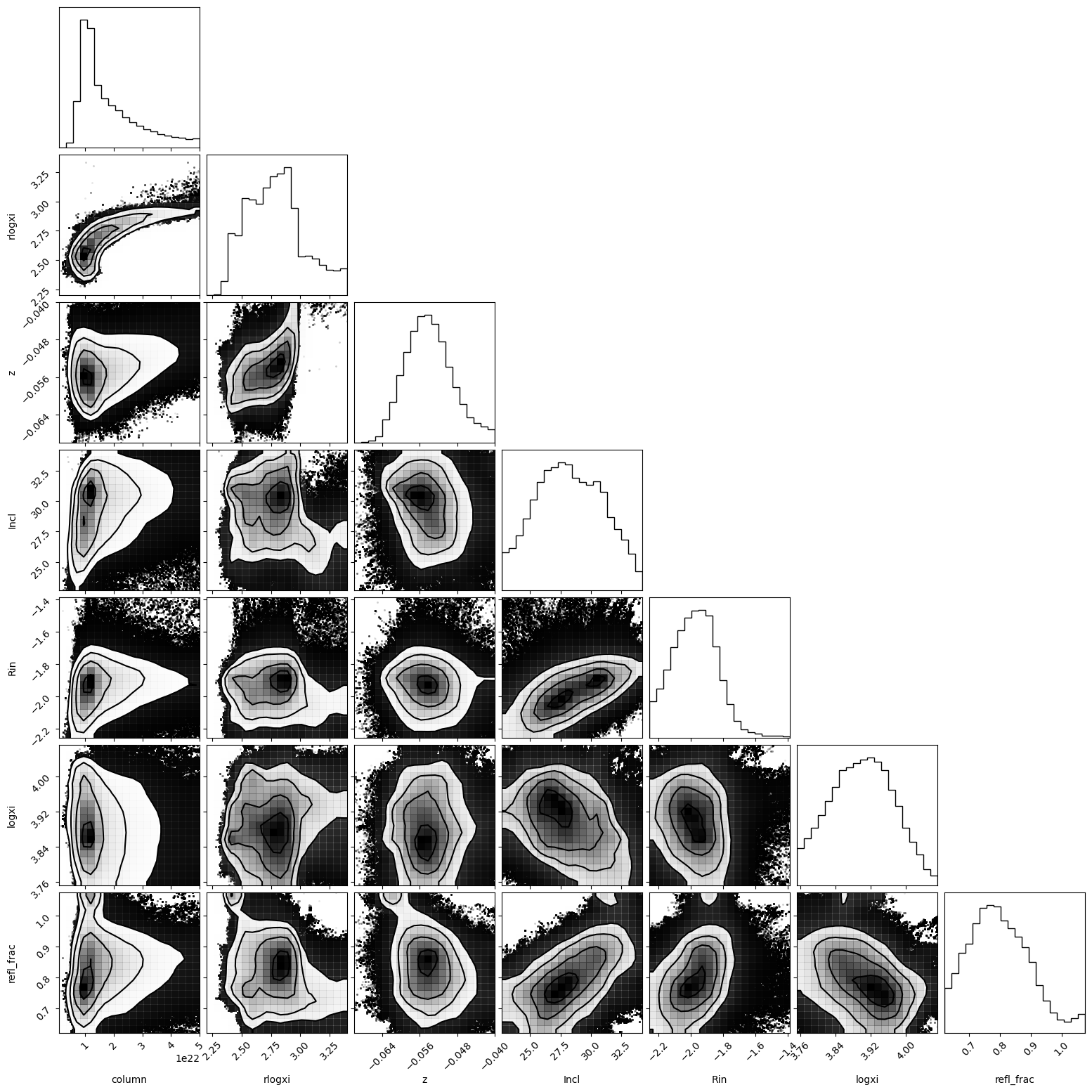}
     \caption{The contour diagram of the different fitting parameters for the 406 spectra with the model 4 generated by the MCMC method. There is no apparent parameter degeneration between these parameters.}
     \label{contour}
\end{figure*}

To derive the physical parameters of the winds, we used xstar v 2.58e code \citep{kallman2001photoionization,kallman2004photoionization} into the spectral fittings, determining the physical parameters of column density, ionization state and outflow velocity. Thus, the constitutions of Models 3 \& 4 are:
$$
    xstar\times tbabs\times(diskbb+relxill) ({\rm Model 3});
$$
$$
    xstar\times tbabs\times(diskbb+relxillCp) ({\rm Model 4}) .
$$
We use the MCMC from xspec to fit the data by these two models, and we have set the turbulent velocity as 6000 $\rm km/s$ for the 406 spectra of 150 eV line width, 10000 $\rm km/s$ for 404 spectra of 250 eV line width and the wind density as $10^{12}\rm cm^{-3}$. Adopting the Goodman-Weare algorithm, we set 40 walkers to run $7\times 10^{6}$ steps and burn the first $1\times 10^{6}$ steps. We test the chain with the integrated auto-correlation time $\tau_{f}$ \citep{foreman2013emcee} and the steps over $5\tau_{f}$ to ensure that the chain has converge (all the $\tau_{f}$ below the $1.5\times 10^{5}$ steps). The model 3 \& 4 fitting results are shown in the model 3 column and the model 4 column of Table \ref{404p} \& \ref{406p}. The 406 spectra and residuals are shown in Fig. \ref{406M4cb}. The example contour diagram of the different parameters produced by MCMC fittings is also displayed in Fig. \ref{contour}.
        
For avoiding the instrumental artifacts, we run the null hypothesis procedure to make the test \citep{tombesi2010evidence}. Firstly, we use the $fakeit$ in xspec to simulate 1000 sets of observations with the same observation exposure time of the 404 and 406 observations and NuSTAR detector response files by Model 2. To check the instrumental artifacts, we add a gaussian absorption with a line centroid energy randomly between 6 keV and 9 keV and freeze the line width to the best-fit value. In the real data, the xstar improves the Chi-stat of $\sim 25$ for ObsID 404, and $\sim 60$ for ObsID 406, so we set the thresholds as 25 for ObsID 404, and 60 for ObsID 406, and we count the number of simulated spectra which improve less Chi-stat than the thresholds. Finally, the confidence is given by the ratio between the counted number of the simulated spectra and the total number of simulated spectra (e.g., 1000), thus, the absorption lines confidence are $\sim 78\%$ for 404 spectra, $\sim 92\%$ for 406 spectra.
	   

\begin{figure*}
     \centering
     \includegraphics[scale=0.7]{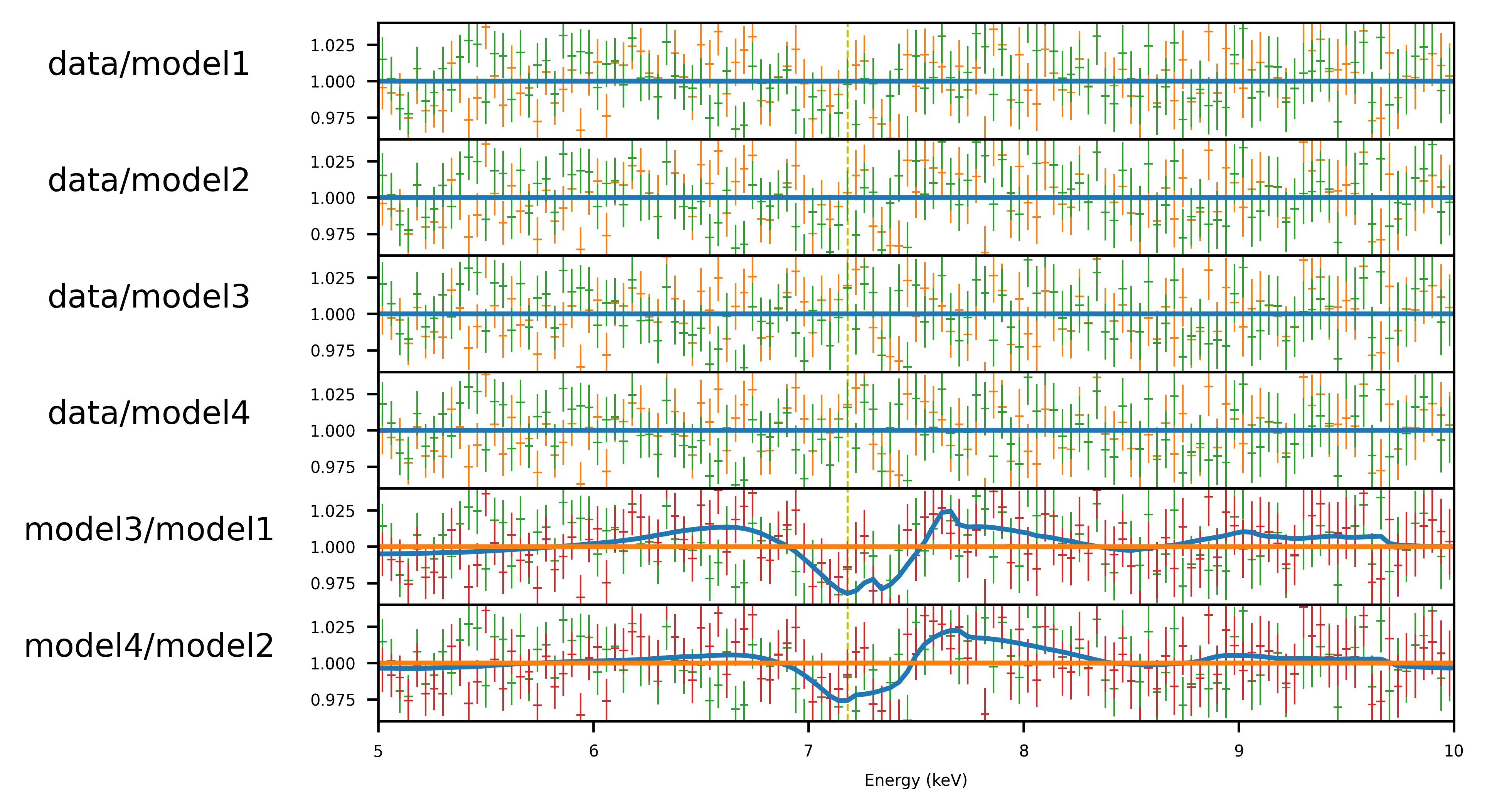}
     \includegraphics[scale=0.7]{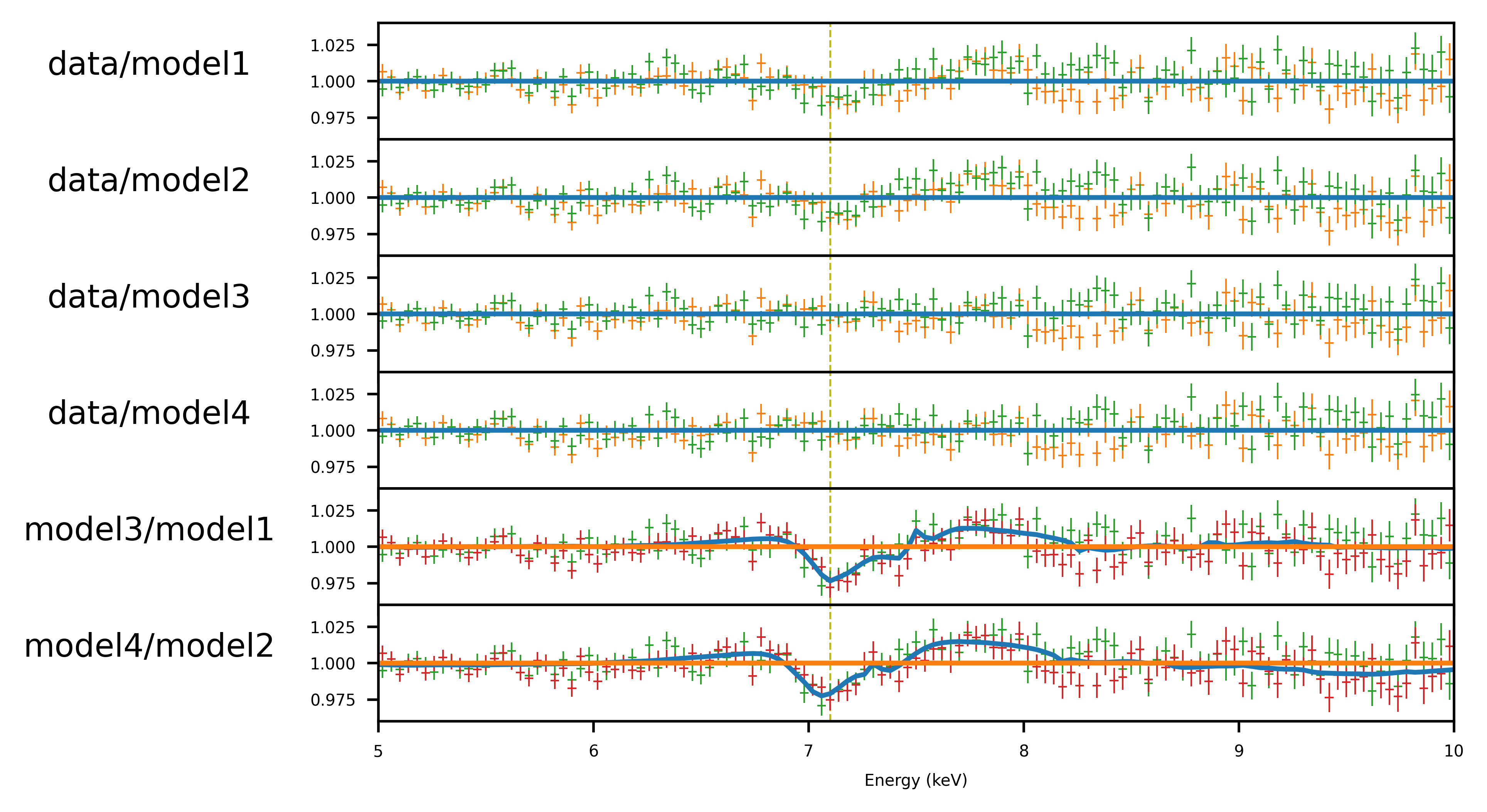}
     \caption{\textbf{Top panel:} This figure illustrates the ratio between different models for the observed spectra of ObsID 404, green for FPMA, orange or red for FPMB. From top to bottom the ratio between data and Model 1, the ratio between data and Model 2, the ratio between data and model 3, the ratio between data and model 4. The last two panels also represent the model ratio of Model 1/Model 3 and Model 3/Model 4 (the blue lines correspond to Model 3 \& 4, which are also illustrated as labels at the left side of the figure). \textbf{Bottom panel:} This is the case of ObsID 406 as the same order as the top panel for ObsID 404.}
     \label{404compare;406compare}
\end{figure*}

In the end, we compared four models and illustrated the significance of the absorption line features in Fig. \ref{404compare;406compare} for the 404 and 406 spectra. From the model data ratios of model3/model1 and model4/model2, we can clearly find the Fe \uppercase\expandafter{\romannumeral26} absorption in both observations. Since the exposure time for ObsID 404 is much shorter than that of ObsID 406, the error bars of the data points are larger for the case of ObsID 404. 

\section{Conclusion and Discussion}\label{conclusion}

In this work, we find that there exist the fast outflows in low hard state and hard intermediate state of MAXI J1348-630 during the 2019 outburst. The photoionization model constrains the parameters of the disk wind with the velocity of the outflows reaching $\sim 10^{4}\rm km s^{-1}$ or even higher, the wind ionization parameter $\xi_{wind}\sim 10^{3}$ $\rm erg\ cm\ s^{-1}$, and the column density getting $(2-20) \times 10^{22}\rm cm^{-2}$. In addition, by fitting the reflection component in the hard X-ray spectrum, we also determine the inclination angle of the binary system to be $\sim 25^\circ -35^\circ$.

The outflows originating from the disks are generally observed in the soft states in different BHXRBs \cite{lee2002high,miller2015powerful,neilsen2018persistent}, but rare in the hard state or hard intermediate state. What mechanisms driving this fast outflows/disk wind are quite uncertain at present. There are three mechanisms producing the disk winds, namely thermal driving, radiation driving and magnetic driving. Our observation results (e.g., based on Models 3 \& 4) demonstrate the consistent values of ionized absorption parameters, which could help to discuss the physical origin of the winds.

With the outflow velocity of over $10^{4}\rm km/s$, we can estimate the production region of disk wind as $r_{o} \ge 200(\frac{v_{e}}{ 10^{4}{\rm km/s}})^{-2}\rm r_{g}$, by assuming that the outflow velocity of the wind is larger than the local escape velocity, where the $v_{e}=\sqrt{\frac{GM_{BH}}{r_{o}}}$ is the escape velocity. The wind production site for MAXI J1348-630 will be located at the inner accretion disk, which is not consistent with the thermal driven winds which predict the winds coming from the larger disk radius of up to $10^4-10^5 \ r_g$ with a typical velocity $\sim 300\rm km/s$ \citep{done2018thermal}. In addition, the multi-temperature black body spectra (the model diskbb) give the disk temperature $\sim 0.65\ \rm keV$, so that the heated surface layer is expelled at the sound speed $\sim 3\times 10^{3}(\frac{E}{0.65\rm keV})^{0.5}\rm km/s$, then in the case of thermal driven winds, this sound speed is higher than the escape velocity, which could drive the outflows with this sound speed at the disk radius outer than $3\times10^{3} r_{g}$. Thus, the thermal driven wind mechanism is not suitable for the fast outflow scenario for MAXI J1348-630. 

Since the radiation pressure would depend mostly on UV absorption lines, then the radiation driven winds will require the wind ionization parameter significantly below a critical value of $10^{3}$ $\rm erg\ cm\ s^{-1}$\citep{proga2002role,king2012extreme,proga2000dynamics}. In the xstar fittings, we found $\xi_{wind}\sim 10^{2.8}-10^{3.3}$ $\rm erg\ cm\ s^{-1}$ in two observations which are around the critical value $\sim 10^{3}$ $\rm erg\ cm\ s^{-1}$\citep{proga2002role,king2012extreme,proga2000dynamics} (we also noted $\xi_{disk}\sim 10^{3.5}-10^{4.5}$ $\rm erg\ cm\ s^{-1}$ for the inner disk which are obtained by fitting the reflection component using the relxill(Cp) model), then the radiation pressure could not be the dominant mechanism to launch the outflow. The magnetic pressure could be dominant near the inner accretion disk around the central black hole \cite{wang2022magnetically,ohsuga2013outflow,cui2020large,fukumura2021modeling,waters2018magnetothermal,fukumura2015magnetically,fukumura2018magnetized}, and the magnetic driven mechanism could launch the fast outflows with a velocity of $> 10^4\ \rm km/s$ within the disk radius of $600 r_g$, and the wind ionization parameter can be in a wide value range of $10^{2.5}-10^{5.5}$ $\rm erg\ cm\ s^{-1}$, the column density of $10^{22}-10^{24}\rm cm^{-2}$, the inclination angle below $\sim 45^\circ$ \cite{wang2022magnetically}. Therefore, the fast disk winds from MAXI J1348-630 in the hard and hard intermediate states would be driven dominantly by magnetic pressure at the inner accretion disk near the black hole.

\section*{Acknowledgements}
We are grateful to the referee for the useful comments to improve the manuscript. This work is supported by the National Key Research and Development Program of China (Grants No. 2021YFA0718500, 2021YFA0718503), the NSFC (12133007, U1838103).


 \bibliographystyle{elsarticle-num} 
 \bibliography{cas-refs}





\end{document}